\documentclass[aps,twocolumn,showpacs,prl,10pt,superscriptaddress,preprintnumbers,nofootinbib]{revtex4-2}
\usepackage{amsmath, amssymb}
\usepackage{physics, bm}
\usepackage{graphicx}
\usepackage{xcolor}
\usepackage[colorlinks, linkcolor=red, citecolor=red, urlcolor=magenta]{hyperref}

\usepackage[normalem]{ulem}
\usepackage{soul}
\usepackage{multirow}
\usepackage{orcidlink}

\newcommand{\beq}[1][]{\begin{equation}\label{#1}}
\newcommand{\eeq}{\end{equation}}
\newcommand{\bea}{\begin{eqnarray}}
\newcommand{\eea}{\end{eqnarray}}
\newcommand{\nn}{\nonumber}

\begin{document}

\preprint{
	{\vbox {
		\hbox{\bf CPTNP-2026-022}
}}}
\vspace*{0.2cm}

\title{Probing Gluon Linear Polarization with Dihadron Fragmentation in $\chi_b$ Decays}

\author{Zhi-Guo He\, \orcidlink{0000-0001-9887-2058}}
\email{zhiguo.he@buct.edu.cn}
\affiliation{Department of Physics and Electronics, School of Mathematics and Physics, Beijing University of Chemical Technology, Beijing 100029, China}

\author{Guanghui Li\,\orcidlink{0009-0001-4822-3321}}
\email{ghli@ihep.ac.cn}
\affiliation{Institute of High Energy Physics, Chinese Academy of Sciences, Beijing 100049, China}
\affiliation{School of Physical Sciences, University of Chinese Academy of Sciences, Beijing 100049, China}

\author{Yang Liu\,\orcidlink{0009-0004-4499-0875}}
\email{liuy01@ihep.ac.cn}
\affiliation{Institute of High Energy Physics, Chinese Academy of Sciences, Beijing 100049, China}
\affiliation{School of Physical Sciences, University of Chinese Academy of Sciences, Beijing 100049, China}

\author{Yu-Jie Tian\,\orcidlink{0009-0002-7071-0916}}
\email{yjtian@ihep.ac.cn}
\affiliation{Institute of High Energy Physics, Chinese Academy of Sciences, Beijing 100049, China}
\affiliation{School of Physical Sciences, University of Chinese Academy of Sciences, Beijing 100049, China}

\author{Xin-Kai Wen\,\orcidlink{0009-0008-2443-5320}}
\email{xinkaiwen@ihep.ac.cn (corresponding author)}
\affiliation{Institute of High Energy Physics, Chinese Academy of Sciences, Beijing 100049, China}
\affiliation{China Center of Advanced Science and Technology, Beijing 100190, China}

\author{Bin Yan\,\orcidlink{0000-0001-7515-6649}}
\email{yanbin@ihep.ac.cn (corresponding author)}
\affiliation{Institute of High Energy Physics, Chinese Academy of Sciences, Beijing 100049, China}
\affiliation{Center for High Energy Physics, Peking University, Beijing 100871, China}

\date{\today}
\begin{abstract}
The dihadron fragmentation function (DiFF) of a linearly polarized gluon has not yet been accessed experimentally, leaving an important aspect of spin-dependent gluon hadronization unexplored. We show that, at leading order, the color-singlet decay channel of the $P$-wave bottomonium state $\chi_{b0}$  produces two energetic gluons with correlated linear polarizations. Within collinear factorization, their fragmentation into separate dihadron pairs generates an Artru--Collins-type angular correlation that provides the first direct probe of the linearly polarized gluon DiFF, while the corresponding semi-inclusive decay rate constrains the unpolarized gluon DiFF.  A spectator-model benchmark indicates percent-level asymmetries, potentially within reach of existing Belle data. A dedicated Belle~II data set would substantially improve the statistical precision, enabling more stringent constraints on the kinematic dependence of the linearly polarized gluon DiFF.
\end{abstract}

\maketitle

\emph{Introduction.---}
Color confinement in quantum chromodynamics (QCD) requires quarks and gluons to hadronize into color-neutral hadrons. Fragmentation functions (FFs) provide a theoretical description of this intrinsically nonperturbative process within QCD factorization, characterizing the hadronization of a parton into specified hadronic final states~\cite{Berman:1971xz,Field:1977fa,Feynman:1978dt}. As universal nonperturbative functions, they provide unique probes of hadronization dynamics in high-energy processes. Spin-dependent FFs further encode spin--orbit correlations, relating parton polarization to measurable momentum and angular correlations among the produced hadrons~\cite{Metz:2016swz}. They therefore provide indispensable tools for unraveling the spin structure and nonperturbative dynamics of hadronization.

Over the past decades, spin-dependent fragmentation has emerged as a powerful framework for probing quark spin structure through experimentally accessible polarization observables. In particular, the chiral-odd Collins FF~\cite{Collins:1992kk,Kang:2015msa,Metz:2016swz,Zeng:2023nnb,Boussarie:2023izj,Zeng:2024gun} and interference DiFFs~\cite{Collins:1993kq,Jaffe:1997hf,Jaffe:1997pv,Bianconi:1999cd,Barone:2001sp,Bacchetta:2011ip,Courtoy:2012ry,Metz:2016swz,Pitonyak:2023gjx,Cocuzza:2023oam,Cocuzza:2023vqs,Rogers:2024nhb,Mahaut:2025hie} have provided access to transverse quark polarization through spin-dependent hadronic angular correlations measured in semi-inclusive deep-inelastic scattering and electron--positron annihilation~\cite{Artru:1995zu,Boer:1997mf,Bacchetta:2003vn,Boer:2003ya,Bacchetta:2008wb,Zhou:2011ba,Pitonyak:2013dsu,Wen:2024cfu,Wen:2024nff,Cheng:2025cuv,Cao:2025qua,Cao:2025wfg,Kang:2025zto,He:2026pxa}. Extending these studies to gluon fragmentation, especially to the fragmentation of linearly polarized gluons, remains challenging. The main difficulty is the absence of sufficiently clean processes that produce gluons with well-controlled polarization correlations while allowing their subsequent hadronization to be isolated. Conventional transverse-momentum-dependent (TMD) observables are further complicated by polarization dilution from TMD evolution and contamination from soft-gluon radiation~\cite{Hatta:2020bgy,Hatta:2021jcd}. Energy correlators have recently been proposed as alternative probes of gluon linear polarization in hard processes~\cite{Song:2025bdj}, but they do not directly determine the underlying nonperturbative gluon FFs. Consequently, spin-dependent gluon fragmentation remains poorly constrained experimentally, leaving gluon hadronization far less understood than its quark counterpart.

To overcome this limitation, we propose in this Letter a novel strategy to directly probe gluon linear polarization through dihadron fragmentation in bottomonium decays at lepton colliders. In particular, the $S$-wave $\eta_b$ and $P$-wave $\chi_{b0,2}$ states serve as clean, energetic gluon sources, as their decays are dominated by the two-gluon channel~\cite{QuarkoniumWorkingGroup:2004kpm}. The large bottom-quark mass introduces a perturbative hard scale, allowing the application of nonrelativistic QCD (NRQCD) factorization to systematically separate the short-distance annihilation dynamics from the nonperturbative long-distance matrix elements (LDMEs) ~\cite{Bodwin:1994jh}.
Implementing NRQCD factorization for the bottomonium decay together with collinear factorization for the subsequent gluon fragmentation, we demonstrate that Artru--Collins-type angular correlations between two dihadron pairs in $\eta_{b}$ and $\chi_{b0}$ decays are sensitive to the linearly polarized gluon DiFF, whereas $\chi_{b2}$ decays do not generate such correlations after summing over its polarizations. This framework yields the first direct probe of gluon linear polarization through dihadron fragmentation.
Additionally, the corresponding semi-inclusive decay rates also provide access to the unpolarized gluon DiFF $D_1^g$, which remains poorly constrained experimentally.
Current knowledge of $D_1^g$ relies either on indirect extractions based on DGLAP evolution of PYTHIA-generated pseudodata~\cite{Cocuzza:2023vqs} or on the assumption that its invariant-mass dependence follows that of the up-quark DiFF~\cite{Mahaut:2025hie}. However, a neural-network analysis without these assumptions indicates that current Belle data possess negligible sensitivity to $D_1^g$~\cite{Mahaut:2025hie}. Our approach thus provides a new avenue for accessing gluon DiFFs and probing gluon linear polarization via spin-dependent hadronization. 

\vspace{3mm}
\emph{Gluon Linear Polarization from $\chi_{b0}$ Decay.---}
%
\begin{figure}
	\centering
	\includegraphics[width=0.9\linewidth]{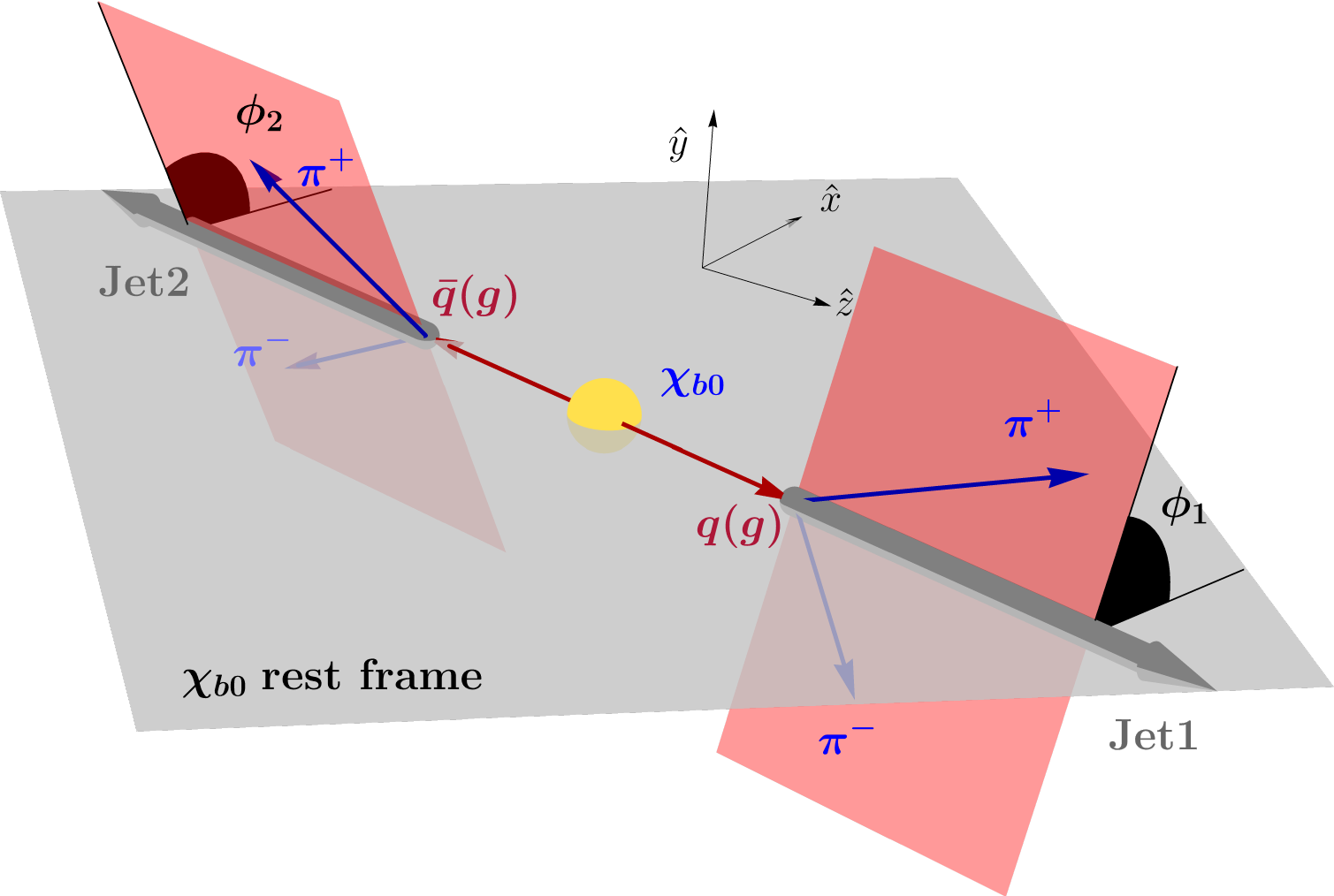}
	\caption{Leading-order kinematic configuration for the production of two $\pi^+\pi^-$ pairs in $\chi_{b0}$ decay, shown in the $\chi_{b0}$ rest frame.}
	\label{fig:geometry}
\end{figure}
%
Given the limited statistics available for $\eta_b$, we focus on the decay channel of $\chi_{b0}$, which can be produced at lepton colliders through the radiative transition $\Upsilon(2S)\to\gamma\chi_{b0}$. For $P$-wave heavy-quarkonium decay, both the color-singlet (CS) $b\bar b({}^3P_0^{[1]})$ and color-octet (CO) $b\bar b({}^3S_1^{[8]})$ configurations contribute at leading order (LO) in the NRQCD velocity expansion~\cite{Bodwin:1994jh}. However, their corresponding short-distance coefficients are associated with distinct subprocesses. At LO in $\alpha_s$, the CS component annihilates into two gluons, whereas the CO component produces a light-quark pair, leading to the fragmentation processes $\chi_{b0}\to gg/q\bar q\to(\pi^+\pi^-)+(\pi^+\pi^-)+X$. The dominant gluon contribution provides a clean probe of gluon dihadron fragmentation, particularly that of linearly polarized gluons.

The decay kinematics are analyzed in the $\chi_{b0}$ rest frame, as illustrated in Fig.~\ref{fig:geometry}. The $\hat z$ axis is chosen along the thrust direction reconstructed from the final-state hadrons, which serves as an experimental approximation to the partonic axis. The two $\pi^+\pi^-$ pairs are selected from opposite hemispheres in events with a large thrust value to suppress multijet contamination~\cite{Belle:2011cur}. For each pion pair, we define the total and relative momenta as $P_i=p_i^{\pi^+}+p_i^{\pi^-}$, and $R_i=(p_i^{\pi^+}-p_i^{\pi^-})/2$, with $i=1,2$. The corresponding invariant mass is $M_i^2=P_i^2$, and $\phi_i$ denotes the azimuthal angle of the transverse component $\bm R_{iT}$ with respect to the thrust axis. We further define the longitudinal momentum fraction as $z_i=2P_i^0/m_\chi$, where $m_\chi$ denotes the mass of the $\chi_{b0}$. Within collinear factorization and for $|\bm P_i|\gg M_i$~\cite{Collins:1993kq,Collins:2011zzd} , the differential distribution of $\pi^+\pi^-$ pairs in $\chi_{b0}$ decay is expressed in terms of $z_i$, $M_i$, and $\phi_i$ in the $\chi_{b0}$ rest frame as
\begin{align}
	&\frac{d\sigma}{\sigma_0 \, dz_1 \, dz_2 \, dM_1 \, dM_2 \, d\phi_1 \, d\phi_2 }\nn\\
	&=H_1^{b} \mathcal{C}_g \, \big[ D_1^g(z_1,M_1)D_1^{g}(z_2,M_2) \nn\\
	&+H_1^{\sphericalangle,g}(z_1,M_1)H_1^{\sphericalangle,g}(z_2,M_2)\cos(2\phi_1-2\phi_2) \big]\nn\\
	&+ H_8^{b} \sum_q \mathcal{C}_q \, D_1^q(z_1,M_1)D_1^{\bar q}(z_2,M_2),
	\label{eq:fac}
\end{align}
where $\sigma_0$ denotes the unpolarized production cross section of $\chi_{b0}$ at lepton colliders. The hard coefficients $\mathcal{C}_{q/g}$ are determined by the corresponding partonic decay rates of $b\bar{b}(^3S_1^{[8]})\to q\bar q$ and $b\bar{b}(^3P_{0}^{[1]})\to gg$, respectively, 
\begin{align}
	\mathcal{C}_q&=\frac{4\pi\alpha_s^2}{3m_{\chi}^2 \Gamma_\chi},&
    \mathcal{C}_g&=\frac{64\pi\alpha_s^2}{3m_{\chi}^4 \Gamma_\chi},
\end{align}
where $\Gamma_\chi$ denotes the total width of the $\chi_{b0}$. The nonperturbative LDMEs for the CS and CO channels are given by
$H_1^{b}=\langle \chi_{b0} | \mathcal{O}({}^3P^{[1]}_0) | \chi_{b0} \rangle$ and $H_8^{b}(\mu_\Lambda)=\langle \chi_{b0} | \mathcal{O}({}^3S^{[8]}_1,\mu_\Lambda) | \chi_{b0} \rangle$,
respectively, with $\mu_\Lambda=m_b$ being the NRQCD factorization scale.

The hadronization process is described by the DiFFs, which encode the probability for a parton to fragment into a specified hadron pair~\cite{Artru:1995zu,Jaffe:1997hf,Jaffe:1997pv,Bianconi:1999cd,Bianconi:1999uc,Barone:2001sp,Boer:2003ya,Bacchetta:2003vn,Bacchetta:2008wb,Bacchetta:2011ip,Zhou:2011ba,Courtoy:2012ry,Radici:2015mwa,Pitonyak:2023gjx,Cocuzza:2023oam,Cocuzza:2023vqs,Rogers:2024nhb,Pitonyak:2025lin,Mahaut:2025hie,Rogers:2026jca}. The unpolarized gluon DiFF $D_1^g$ governs the azimuthally independent production rate, while the linearly polarized gluon DiFF $H_1^{\sphericalangle,g}$ generates the characteristic $\cos(2\phi_1-2\phi_2)$ modulation. The unpolarized quark DiFF $D_1^q$ contributes only to the azimuthally independent rate and dilutes, but cannot generate, the Artru--Collins asymmetry, since the scalar $\chi_{b0}$ decay does not induce a transverse-spin correlation between the produced quark and antiquark. The scale arguments of the DiFFs are suppressed, and the factorization scale is fixed at $\mu=m_\chi$. The gluon-induced Artru--Collins-type asymmetry can then be isolated through the weighted average of $\cos(2\phi_1-2\phi_2)$,
\begin{widetext}
	\begin{align}
		A_{12}\equiv 2 \left \langle \cos(2\phi_1-2\phi_2)\right\rangle
		=\frac{m_b^2\mathcal{C}_g H_1^{\sphericalangle,g}(z_1,M_1)H_1^{\sphericalangle,g}(z_2,M_2)}{\rho_8(m_b)\sum_q\mathcal{C}_qD_1^q(z_1,M_1)D_1^{\bar q}(z_2,M_2)+m_b^2\mathcal{C}_gD_1^g(z_1,M_1)D_1^g(z_2,M_2)},
		\label{eq:A12}
	\end{align}
\end{widetext}
where we introduce the ratio $\rho_8(m_b)=H_8^{b}(m_b) m_b^2 / H_1^{b}$ between LDMEs of CO $H_8^{b}$ and CS $H_1^{b}$ for convenience. This ratio was extracted from the CLEO measurement of the inclusive decay $\chi_{bJ}(1P)\to D^0X$~\cite{CLEO:2008bsq} and independently obtained from lattice NRQCD calculations~\cite{Bodwin:2001mk,Bodwin:2007zf}, with the two results showing a notable discrepancy. We recently showed that dihadron fragmentation observables in $\chi_{b2}$ decays provide an independent determination of this ratio, offering a promising avenue for clarifying this discrepancy~\cite{He:2026pxa}. 

\begin{figure}
	\centering
	\includegraphics[width=0.9\linewidth]{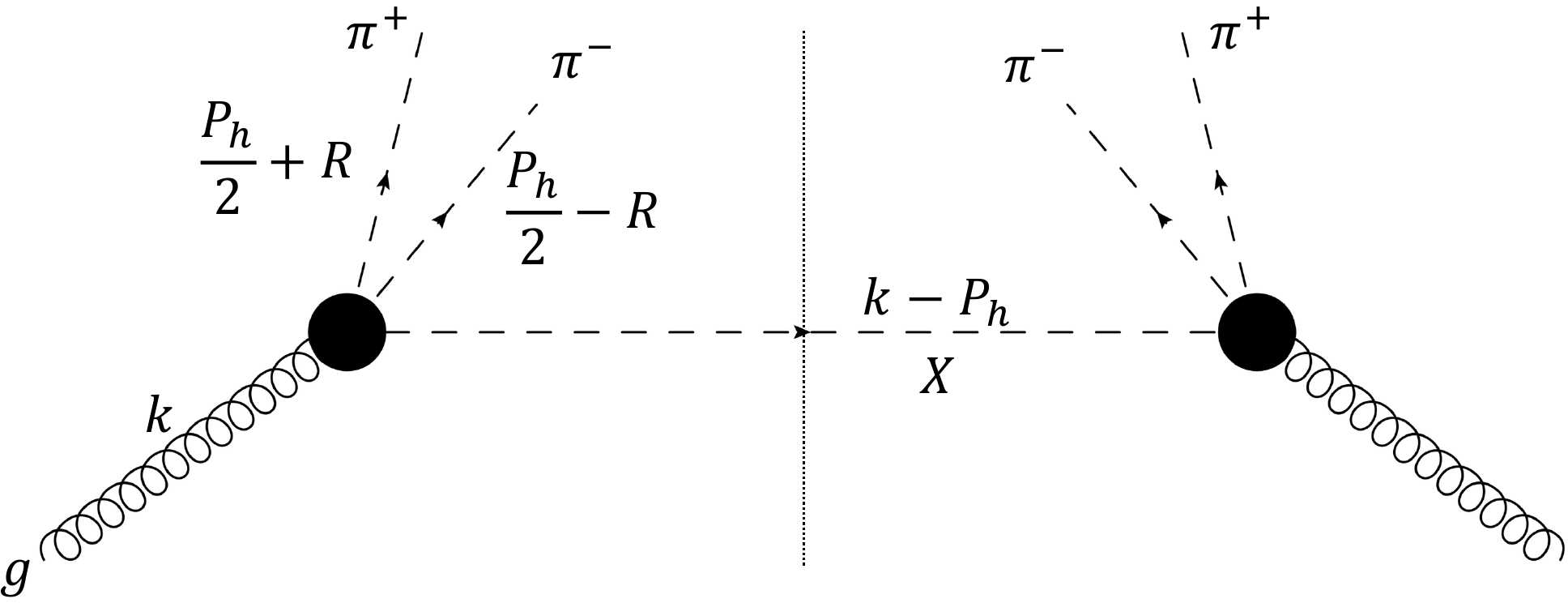}
	\caption{Schematic diagram of the gluon fragmentation correlator $\Delta^{g,ij}$ for $g\to\pi^+\pi^-X$ in the scalar-spectator model.}
	\label{fig:gluon_spec}
\end{figure}

\vspace{3mm}
\emph{Spectator Model for Gluon Fragmentation.---}
To estimate the Artru--Collins-type asymmetry induced by linearly polarized gluons, we employ a spectator model for the nonperturbative fragmentation process $g\to\pi^+\pi^-X$, in which the unobserved remnant system is approximated by an effective on-shell spectator state~\cite{Jakob:1997wg,Bacchetta:2006un,Xie:2022lra}. For simplicity, we restrict our analysis to a scalar spectator with momentum $P_s$ and mass $M_s$, as illustrated in Fig.~\ref{fig:gluon_spec}. The corresponding operator-defined fragmentation correlator is modeled through a gluon--hadron--spectator vertex. The pion-pair momentum is given by $P_h=k-P_s$, where $k$ is the momentum of the fragmenting gluon, with $M_h^2=P_h^2$ and $R$ denoting the relative momentum of the two pions. The unintegrated gluon fragmentation correlator then takes the form
\begin{align}\label{eq:Delta-def}
	\Delta^{g,ij}(k,P_h,R)
	&= \frac{1}{N_c^2-1}\frac{1}{(2\pi)^4} 2\pi\,\delta\!\left((k-P_h)^2-M_s^2\right)
	\nn\\
	&\times
	(k^-)^2G^{j\nu *}_{ab'}(k)\Gamma^{*}_{\nu,b'c}
	G^{i\mu}_{ab}(k)\Gamma_{\mu,bc},
\end{align}
where $1/(N_c^2-1)$ accounts for the initial gluon color average, $i,j$ denote transverse gluon polarization indices, and $a, b, c$ are color indices. The gluon propagator $G^{i\mu}_{ab}(k)$ is taken in the light-cone gauge. 

The effective vertex is parameterized as
\begin{align}\label{eq:vertex}
	\Gamma_{ab}^{\mu}(k,P_h,R)
	=
	\delta_{ab}\,k^2
	\left[
	e^{-k^2/\Lambda_s^2}F_sV_s^\mu
	+
	e^{-k^2/\Lambda_p^2}F_{pi}V_{pi}^\mu
	\right],
\end{align}
where $F_{s,pi}$ denote the invariant-mass-dependent form factors and $V_{s,pi}^{\mu}$ represent the Lorentz structures for the $s$- and $p$-wave components. 
The factors $k^2\exp(-k^2/\Lambda_{s,p}^2)$ suppress the low- and high-virtuality regions, thereby concentrating the fragmentation strength at intermediate timelike virtualities characterized by $\Lambda_{s,p}$.

The effective structures are constructed by treating the $s$- and $p$-wave $\pi^+\pi^-$ states as effective scalar and vector degrees of freedom, respectively, and are constrained by Lorentz covariance and gauge invariance:
\begin{align}
	V_s^\mu &=
	(k\!\cdot\!P_h)P_s^\mu-(k\!\cdot\!P_s)P_h^\mu,
	\nn\\
	V_{p1}^\mu &=
	(k\!\cdot\!R)P_s^\mu-(k\!\cdot\!P_s)R^\mu,
	\nn\\
	V_{p2}^\mu &=
	(k\!\cdot\!R)P_h^\mu-(k\!\cdot\!P_h)R^\mu.
\end{align}
Under $R\to-R$, which reverses the relative momentum of the pion pair, $V_s^\mu$ is even, whereas $V_{p1}^\mu$ and $V_{p2}^\mu$ are odd. After truncating the relative partial-wave expansion at $L=1$, these structures are identified with the $s$- and $p$-wave sectors, respectively. All three structures satisfy $k_\mu V_{s,p1,p2}^{\mu}=0$, ensuring the transversality of the gluon coupling.

The $s$-wave amplitude describes the nonresonant $\pi^+\pi^-$ continuum, while the $p$-wave amplitudes are dominated by intermediate vector-meson states, primarily the $\rho$ and $\omega$. The invariant-mass-dependent form factors are parameterized as~\cite{Bacchetta:2006un}
\begin{align}
	F_s&=f_s,\nn\\
	F_{pi}
&=
f_{\rho_i}
\frac{(M_h^2-M_\rho^2)-i\Gamma_\rho M_\rho}
{(M_h^2-M_\rho^2)^2+\Gamma_\rho^2M_\rho^2}
\nn\\
&+
f_{\omega_i}
\frac{(M_h^2-M_\omega^2)-i\Gamma_\omega M_\omega}
{(M_h^2-M_\omega^2)^2+\Gamma_\omega^2M_\omega^2}
\nn\\
&+
if'_{\omega_i}
\frac{
\sqrt{\lambda(M_\omega^2,M_h^2,m_\pi^2)}
\Theta(M_\omega-m_\pi-M_h)}
{4\pi\Gamma_\omega M_\omega^2
\sqrt[4]{4M_\omega^2m_\pi^2+\lambda(M_\omega^2,M_h^2,m_\pi^2)}},
\label{eq:Fp}
\end{align}
where $\Theta$ denotes the unit step function and
\beq
\lambda(M_\omega^2,M_h^2,m_\pi^2)
=
[M_\omega^2-(M_h+m_\pi)^2]
[M_\omega^2-(M_h-m_\pi)^2].
\eeq
Here, $M_\rho$ and $\Gamma_\rho$ ($M_\omega$ and $\Gamma_\omega$) denote the masses and widths of the $\rho$ ($\omega$) meson. The first two terms in $F_{pi}$ describe the resonant channels
$g\to\rho X\to(\pi^+\pi^-)X$ and $g\to\omega X\to(\pi^+\pi^-)X$, respectively. The final term accounts for the contribution of the three-pion decay channel $\omega\to\pi^+\pi^-\pi^0$, with the unobserved $\pi^0$ integrated out.

The collinear gluon fragmentation correlator is obtained by integrating over the unobserved components of the fragmenting gluon momentum
\beq
\Delta^{g,ij}(z,\xi,\bm R_T)=\int d^2{\bm k_T}dk^+\Delta^{g,ij}(k,P_h,R)\bigg|_{k^- = P_h^- / z},
\eeq
where $z=\hat z_1+\hat z_2$ denotes the longitudinal momentum fraction of the dihadron system with respect to the fragmenting gluon, with $\hat z_{1,2}$ the fractions carried by the individual hadrons. The variable $\xi=(\hat z_1-\hat z_2)/z$ characterizes the longitudinal momentum imbalance between the two hadrons within the dihadron system. The gluon DiFFs are then defined through the following projections using the number-density normalization convention adopted by the JAM Collaboration~\cite{Pitonyak:2023gjx}:
\begin{align}
&D_1^g(z,\xi,|\bm R_T|^2)=\mathcal{A}\delta_T^{ij}\Delta^{g,ij}(z,\xi,\bm R_T), &\nn\\
&H_1^{\sphericalangle,g}(z,\xi,|\bm R_T|^2)=\frac{M_h^2}{|\bm R_T|^2}\mathcal{A} T_R^{ij}\Delta^{g,ij}(z,\xi,\bm R_T),
\end{align}
where
\beq
\mathcal{A}=\frac{z^2}{16\pi^3(1-\xi^2)P_h^-},\quad T_R^{ij}=\frac{2R_T^iR_T^j}{|\bm R_T|^2}-\delta_T^{ij}.
\eeq
Here, $\delta_T^{ij}$ contracts the transverse indices of the gluon correlator and extracts the unpolarized component, while the symmetric traceless tensor $T_R^{ij}$ projects the tensor structure associated with linearly polarized gluons.
For comparison with the phenomenological parametrization entering Eq.~\eqref{eq:fac}, the DiFFs are expressed in terms of the $(z,M_h)$-dependent functions,
\begin{align}
&D_1^g(z,M_h)= \frac{M_h}{4}\int d\xi (1-\xi^2)D_1^g(z,\xi,|\bm R_T|^2), &\\
&H_1^{\sphericalangle,g}(z,M_h)= \frac{M_h}{4}\int d\xi \frac{|\bm R_T|^2}{M_h^2}(1-\xi^2)H_1^{\sphericalangle,g}(z,\xi,|\bm R_T|^2).\nn
\end{align}

Using the JAM estimate of the unpolarized gluon DiFF~\cite{Cocuzza:2023vqs} as a benchmark, we fit the scalar-spectator model parameters in the relevant $(z,M_h)$ region, subject to $M_s+M_h<M_\chi/2$:
\begin{align}\label{eq:fit}
	&f_s = 2.27 \times 10^{3}\,\mathrm{GeV}^{-5},
	& &f_{\rho_1}= 8.02 \times 10^{-1}\,\mathrm{GeV}^{-3},
	\nn\\
	&f_{\omega_1}= -9.17 \times 10^{-3}\,\mathrm{GeV}^{-3},
	& &f'_{\omega_1}= -5.97 \,\mathrm{GeV}^{-3},
	\nn\\
	&f_{\rho_2}= -1.74 \times 10^{1}\,\mathrm{GeV}^{-3},
	& &f_{\omega_2}= 1.49 \,\mathrm{GeV}^{-3},
	\nn\\
	&f'_{\omega_2}= 1.01 \times 10^{1}\,\mathrm{GeV}^{-3},
	& &\Lambda_s= 1.83\,\mathrm{GeV},\nn\\
	&\Lambda_p= 3.68\,\mathrm{GeV},
	& &M_s= 2.88\,\mathrm{GeV}.
\end{align}
The JAM gluon DiFF is inferred indirectly through DGLAP evolution and remains only weakly constrained by existing data. Consequently, the fitted spectator-model parameters carry sizable uncertainties. We therefore use the best-fit values solely as a benchmark for estimating the magnitude of the gluon-induced asymmetry. Measurements of dihadron production in $\chi_{b0}$ decays would provide new direct constraints on these poorly known gluon DiFFs.

\begin{figure}
	\centering
	\includegraphics[width=\linewidth]{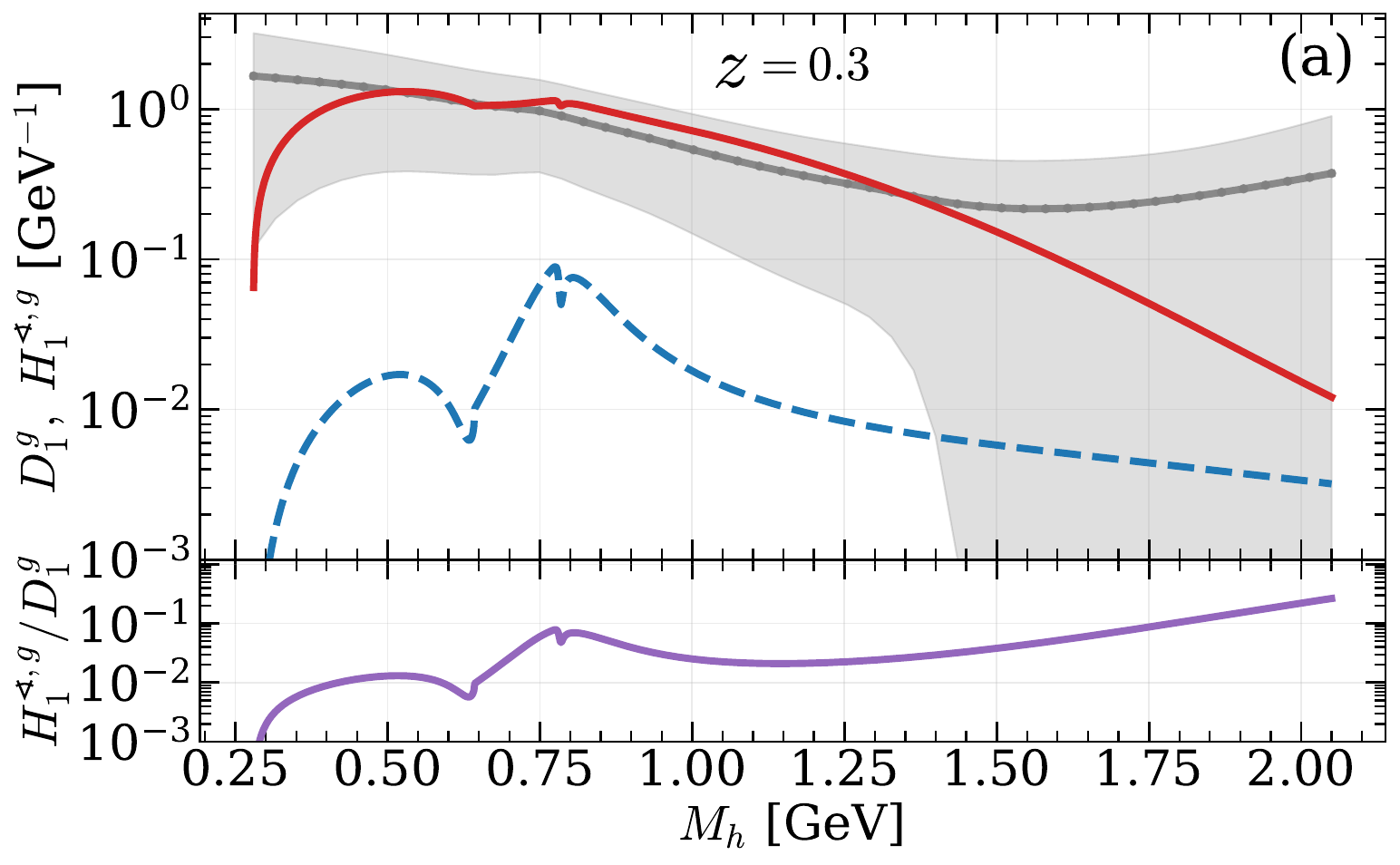}
    \includegraphics[width=\linewidth]{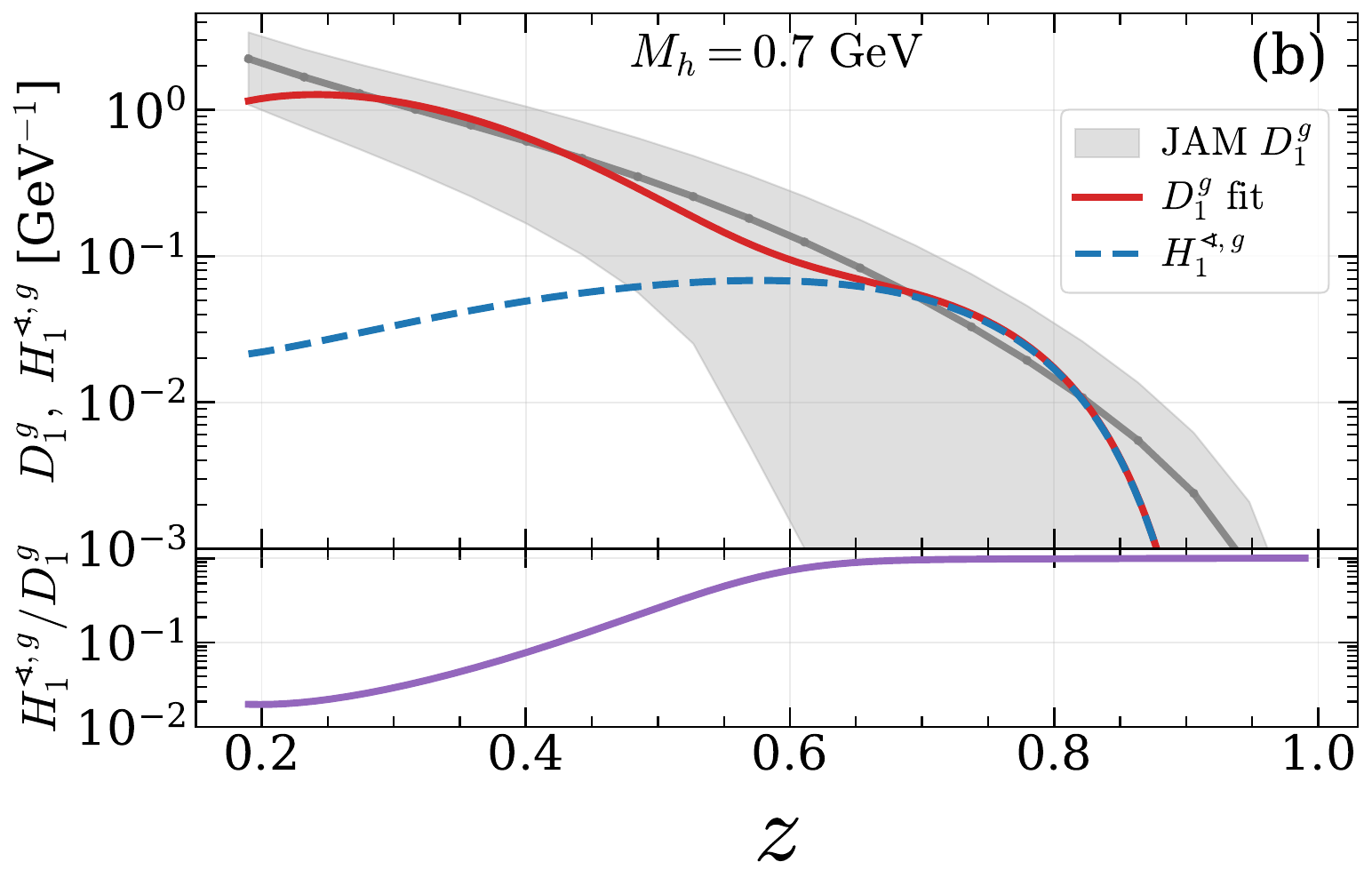}
	\caption{Scalar-spectator model predictions for the gluon DiFFs $D_1^g$ (red) and $H_1^{\sphericalangle,g}$ (blue) using the best-fit parameters in Eq.~\eqref{eq:fit}. The $D_1^g$ result is compared with the available JAM determination (gray)~\cite{Cocuzza:2023vqs}.}
	\label{fig:gluon_DiFFs}
\end{figure}

Figure~\ref{fig:gluon_DiFFs} presents the scalar-spectator model benchmark estimates for the gluon DiFFs $D_1^g$ and $H_1^{\sphericalangle,g}$.
The gray curves and bands denote the central values and $1\sigma$ uncertainties of the unpolarized gluon DiFF $D_1^g$ from the JAM global analysis~\cite{Cocuzza:2023vqs}. The red solid and blue dashed curves show our model results for $D_1^g$ and $H_1^{\sphericalangle,g}$, respectively, obtained with the fitted parameters given above. The predicted DiFFs exhibit characteristic invariant-mass structures associated with the $\rho$- and $\omega$-resonance regions. The ratio $H_1^{\sphericalangle,g}/D_1^g$ characterizes the spin-analyzing power of the dihadron system. Its enhancement in the resonance region reflects the increasing relative contribution of the resonance-driven $p$-wave component compared with the nonresonant $s$-wave continuum contribution.

\vspace{3mm}
\emph{Sensitivity and Discussion.---}
For the numerical evaluation of the Artru--Collins-type asymmetry $A_{12}$ in $\chi_{b0}$ decays, we use the JAM quark DiFFs~\cite{Cocuzza:2023vqs}, while the gluon DiFFs $D_1^g$ and $H_1^{\sphericalangle,g}$ are modeled using the scalar-spectator model described above. To estimate the expected sensitivity of $A_{12}$, we consider only statistical uncertainties and estimate
\begin{equation}
\delta A_{12}=\sqrt{\frac{2-A_{12}^2}{N}}\simeq \sqrt{\frac{2}{N}},
\end{equation}
where $N$ denotes the number of selected events after kinematic cuts, and the approximation holds for $|A_{12}|\ll 1$. 

\begin{figure}
	\centering
	\includegraphics[width=0.9\linewidth]{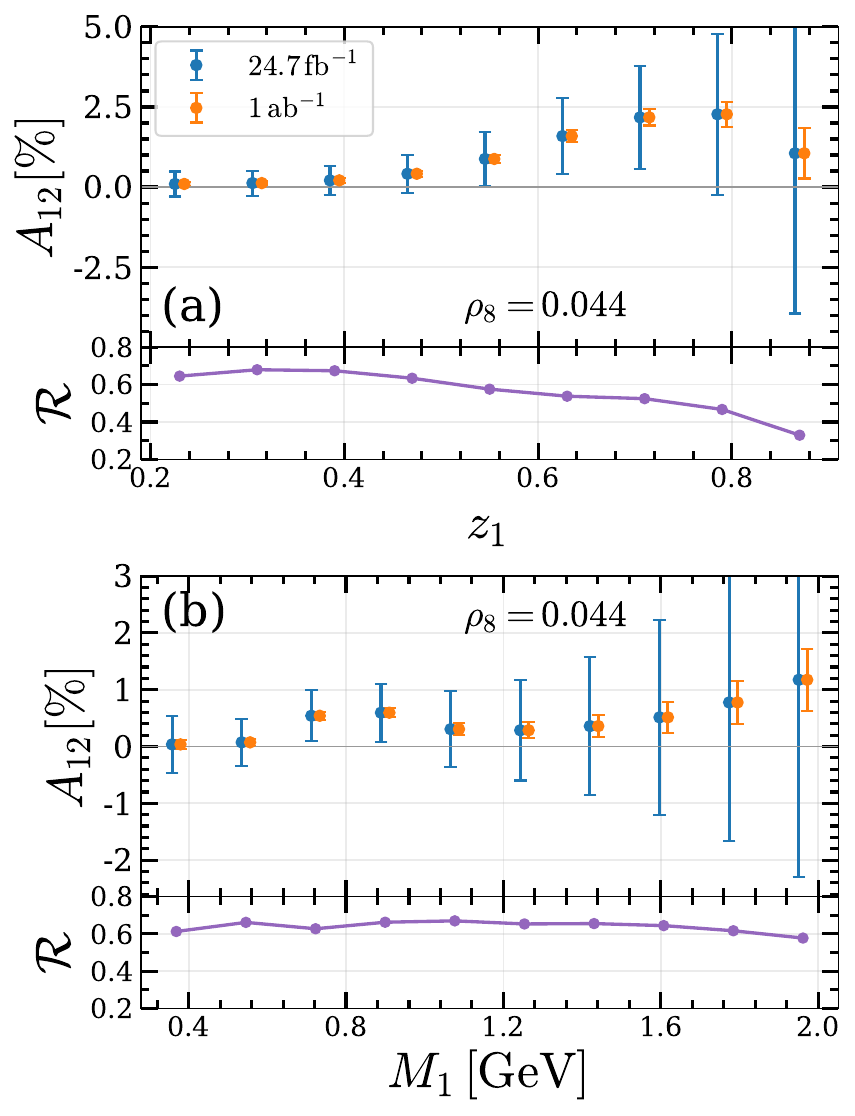}
	\caption{The Artru--Collins-type asymmetry $A_{12}$ (upper panels) and the ratio $\mathcal{R}=A_{12}(\rho_8=0.160)/A_{12}(\rho_8=0.044)$ (lower panels) in $\chi_{b0}$ decays, shown as functions of $z_1$ (a) and $M_1$ (b). The upper panels use $\rho_8=0.044$ from lattice NRQCD~\cite{Bodwin:2007zf}, while the ratio compares this value with the CLEO extraction $\rho_8=0.160$~\cite{CLEO:2008bsq}. The remaining kinematic variables are integrated over, and the uncertainties shown are statistical only.}
	\label{fig:asy}
\end{figure}
Figure~\ref{fig:asy} shows the model prediction for the asymmetry $A_{12}$ as a function of $z_1$ (a) and $M_1$ (b) for the Belle data set collected at the $\Upsilon(2S)$ resonance, corresponding to an integrated luminosity of $\mathcal{L}=24.7~\mathrm{fb}^{-1}$ (blue). The remaining kinematic variables are integrated over the allowed ranges $z_i\in [0.19,0.99]$ and $M_i\in [0.28,2.05]~\mathrm{GeV}$~\cite{Cocuzza:2023vqs}, with $z_i\geq 2M_i/m_\chi$.
We adopt the lattice result $\rho_8(m_b)=0.044$~\cite{Bodwin:2007zf} at $m_b\simeq4.6~\mathrm{GeV}$ as the reference value. The spectator-model benchmark yields percent-level asymmetries, suggesting that existing Belle data may be sensitive to the linearly polarized gluon DiFF $H_1^{\sphericalangle,g}$. Combined with measurements of the $\pi^+\pi^-$ dihadron semi-inclusive decay rates, the same data could also directly constrain the unpolarized gluon DiFF $D_1^g$. Assuming an integrated luminosity of $\mathcal{L}=1~\mathrm{ab}^{-1}$ at Belle II (orange), the statistical precision would be substantially improved, allowing detailed studies of the kinematic dependence of both gluon DiFFs.

We next assess the sensitivity of the predicted asymmetry to the poorly known LDME ratio $\rho_8$. 
The purple lines in Fig.~\ref{fig:asy} show the ratio $\mathcal{R}=A_{12}(\rho_8=0.160)/A_{12}(\rho_8=0.044)$ as a function of $z_1$ (a) and $M_1$ (b), obtained after integrating over the remaining kinematic variables. Here, 
$\rho_8=0.044$~\cite{Bodwin:2007zf} is the lattice NRQCD value adopted above, whereas $\rho_8=0.160$ is extracted from CLEO measurements~\cite{CLEO:2008bsq}. A larger $\rho_8$ enhances the unpolarized quark contribution in the denominator of Eq.~\eqref{eq:A12}, thereby suppressing the asymmetry and yielding $\mathcal{R}<1$. The suppression is weak at small $z_1$, where the two-gluon CS channel dominates, but becomes more pronounced at larger $z_1$ as the unpolarized gluon DiFF falls more rapidly than the quark DiFFs. In contrast, the ratio $\mathcal{R}$ shows only mild dependence on $M_1$. Although the decay rate from the CO light-quark channel is much smaller than that from the CS two-gluon channel, the unpolarized quark DiFF from the JAM analysis is about $4$--$5$ times larger than the gluon DiFF in the relevant kinematic region. Consequently, the quark contribution can still lead to a sizable dilution of the model-estimated asymmetry. This highlights the potential advantage of extending the approach to $\eta_b$ decays, where the absence of the quark channel at LO offers a cleaner probe of gluon fragmentation once sufficient experimental data become available.

\vspace{3mm}
\emph{Conclusion.---}
In this Letter, we have proposed a new strategy to probe gluon linear polarization through dihadron fragmentation in bottomonium decays at lepton colliders. Combining collinear and NRQCD factorization, we showed that the dominant two-gluon decay of $\chi_{b0}$ generates a characteristic $\cos (2\phi_1-2\phi_2)$ correlation between dihadron pairs in opposite hemispheres. The resulting Artru--Collins-type asymmetry provides the first direct probe of the linearly polarized gluon DiFF $H_1^{\sphericalangle,g}$, while the corresponding dihadron semi-inclusive decay rate constrains the unpolarized gluon DiFF $D_1^g$. Using a spectator-model benchmark together with global fits of the quark DiFFs, we obtain percent-level asymmetries, indicating that existing Belle data may already be sensitive to this observable, with substantially improved precision expected at Belle~II. 
The framework can also be extended to $\eta_b$ decays and to other polarization-sensitive observables, including TMD spin correlations and energy correlators. Bottomonium decays therefore provide a new and theoretically controlled avenue for accessing gluon FFs and investigating the role of gluon linear polarization in hadronization.

\vspace{3mm}
We thank Z. Lu  for helpful discussion on the spectator model. This work is partly supported by the National Natural Science Foundation of China under Grant Nos.~12547174, ~12422506, ~12342502 and CAS under Grant No.~E429A6M1, and is partly supported by the China Postdoctoral Science Foundation under Grant No.~2026M793746 and by Fundamental Research Funds for the Central Universities through Grant No. buctrc202432. The authors gratefully acknowledge the valuable discussions and insights provided by the members of the Collaboration on Precision Tests and New Physics (CPTNP).

\bibliographystyle{apsrev}
\bibliography{reference}

\end{document}